\documentstyle[12pt]{article}
\def\singlespace {\smallskipamount=3.75pt plus1pt minus1pt
                  \medskipamount=7.5pt plus2pt minus2pt
                  \bigskipamount=15pt plus4pt minus4pt
                  \normalbaselineskip=15pt plus0pt minus0pt
                  \normallineskip=1pt
                  \normallineskiplimit=0pt
                  \jot=3.75pt
                  {\def\smallskip {\vskip\smallskipamount}}
                  {\def\medskip   {\vskip\medskipamount}}
                  {\def\bigskip   {\vskip\bigskipamount}}
                  {\setbox\strutbox=\hbox{\vrule
                    height10.5pt depth4.5pt width 0pt}}
                  \parskip 7.5pt
                  \normalbaselines}
\def\middlespace {\smallskipamount=5.625pt plus1.5pt minus1.5pt
                  \medskipamount=11.25pt plus3pt minus3pt
                  \bigskipamount=22.5pt plus6pt minus6pt
                  \normalbaselineskip=22.5pt plus0pt minus0pt
                  \normallineskip=1pt
                  \normallineskiplimit=0pt
                  \jot=5.625pt
                  {\def\smallskip {\vskip\smallskipamount}}
                  {\def\medskip   {\vskip\medskipamount}}
                  {\def\bigskip   {\vskip\bigskipamount}}
                  {\setbox\strutbox=\hbox{\vrule
                    height15.75pt depth6.75pt width 0pt}}
                  \parskip 11.25pt
                  \normalbaselines}
\def\doublespace {\smallskipamount=7.5pt plus2pt minus2pt
                  \medskipamount=15pt plus4pt minus4pt
                  \bigskipamount=30pt plus8pt minus8pt
                  \normalbaselineskip=30pt plus0pt minus0pt
                  \normallineskip=2pt
                  \normallineskiplimit=0pt
                  \jot=7.5pt
                  {\def\smallskip {\vskip\smallskipamount}}
                  {\def\medskip   {\vskip\medskipamount}}
                  {\def\bigskip   {\vskip\bigskipamount}}
                  {\setbox\strutbox=\hbox{\vrule
                    height21.0pt depth9.0pt width 0pt}}
                  \parskip 15.0pt
                  \normalbaselines}

\include{dspace12}
\def\be{\begin{equation}}
\def\ee{\end{equation}}
\def\bea{\begin{eqnarray}}
\def\eea{\end{eqnarray}}
\def\nn{\nonumber}
\def\th{\theta}
\def\ph{\phi}
\def\lt{\left}
\def\rt{\right}
\def\sect #1{\setcounter{equation}{0}}

\begin{document}
 \middlespace
\begin{flushleft}
{\Large
{\bf Gravitational and electromagnetic fields of a charged tachyon}
}
\end{flushleft}
\begin{flushleft}
K S VIRBHADRA \footnote[1]{
Present address\ :\ Theoretical Astrophysics Group, Tata Institute
of Fundamental Research, Homi Bhabha Road, Colaba, Bombay
400005,  India; E-mail\ :\  shwetketu@tifrvax.tifr.res.in
} \\
Physical Research Laboratory,
Navrangpura, Ahmedabad 380 009,
India.
\end{flushleft}
\vspace{1.0in}
{\bf Abstract} - An  axially symmetric exact solution of the
Einstein-Maxwell  equations is obtained and is interpreted to give the
gravitational and electromagnetic
fields of a charged tachyon. Switching off the charge parameter yields the
solution for the uncharged tachyon which was earlier obtained by Vaidya.
The null surfaces for the charged tachyon are discussed.\\
\vspace{-0.3in}
\begin{flushleft}
Keywords.\ \  Einstein-Maxwell equations; Tachyon \\
PACS Nos.\ \  04.20; 04.40; 14.80.
\end{flushleft}
\vspace{0.7in}
\begin{center}
( To appear in Pramana - J. Phys. {\bf 45} 1995)
\end{center}

\newpage

\begin{flushleft}
{\large {\bf 1. Introduction}}\\
\end{flushleft}

The tardyons, the luxons, and the tachyons are the names given
to particles which move, respectively, with velocities smaller than, equal to,
and greater than the speed of light in vacuum $[1-2]$. Among
these three classes of objects, tachyons have remained
undetected. However, the subject of the superluminary objects
have been fascinating many physicists' minds. Lucretius (50 B.C.)
was probably the first scientist who mentioned
such objects $[3]$. Few years before   the
special theory of relativity was given, Thomson, Heavyside,
and Sommerfeld had investigated the questions arising from the
assumptions that the superluminary objects exist in the nature $[4]$.
However, with the outset of the special theory of relativity (STR), due
to a misinterpretation by Einstein himself, a wrong idea prevailed
that the non-existence of tachyons is a direct consequence of
the STR $[1,4]$. To this end, Bilaumik et al
 (BDS) $[4]$ reexamined this subject and
found that it is rather the special theory of relativity that
suggests a possibility for existence of superluminary objects.
In order to have the  energy and momentum (which are measurable
quantities) real, BDS $[4]$ hypothesized the tachyons to have their
proper masses imaginary. Likewise, the proper lengths and the proper
times of tachyons are also imaginary quantities. The velocity
of a tachyon increases on the loss of its energy.

The causality problems (raised by Tolman, Schmidt, and Terletski)
due to the assumption that the tachyons exist were resolved by
Sudarshan and his co-workers $[2,3]$.
BDS $[4]$ suggested that the Cerenkov effect is likely to be an avenue for
the detection of the faster-than-light objects if they carry
electric charge.  Many experimentalists
$[5-8]$ have put considerable effort to detect tachyons. Their experiments
could not confirm the existence of tachyons.
 However, inspired by the  Gell-Mann's
totalitarian principle (which states that in physics anything
which is not forbidden is compulsory) $[2,3]$, the subject of
tachyons continued to be of interest to many researchers. Recami $[3]$
presented a review  on this subject.

Narlikar and Sudarshan $[9]$ studied
the propagation of tachyons in an expanding universe.
They showed that the pre-mordial
tachyons in a big-bang universe could not survive unless it had
very large energy in the beginning.
The trajectories of tachyons in the Schwarzschild background
were studied by Narlikar and Dhurandhar $[10]$. These
investigations were further extended by Dhurandhar $[11]$ in the Kerr geometry.
Vaidya $[12]$ obtained a static,
axially symmetric vacuum solution of the Einstein equations and
interpreted it as describing  the gravitational field of a tachyon.
He found that there is a gravitational repulsion between a
tachyon and a tardyon. Raychaudhuri $[13]$ showed that there is
an attraction between tachyons themselves.
We $[14]$ obtained the gravitational field of a tachyon (uncharged) in a de
Sitter universe.
The electromagnetic
interactions between a charged tachyon and a charged tardyon,
and between charged tachyons themselves have not been investigated.
Therefore,  it is  of interest to obtain the gravitational
as well as the  electromagnetic fields of a charged tachyon.
In the present paper, we obtain   an  axially symmetric exact solution
of the Einstein-Maxwell
equations and  interprete it  to give the gravitational and
electromagnetic fields of a charged tachyon.
We use the geometrized units ($G=1, c=1$) and follow the
convention that the Latin indices run from $0$ to $3$.

\begin{flushleft}
{\large {\bf 2. Einstein-Maxwell Equations}}\\
\end{flushleft}

The Einstein-Maxwell equations are
\be
R_i^{\ k} - \frac{1}{2}\ g_i^{\ k} R = 8 \pi \lt( T_i^{\ k} +
E_i^{\ k}\rt) ,
\ee
where
\be
E_i^{\ k} = \frac{1}{4\pi} \lt[-F_{im} F^{km} + \frac{1}{4}\
                              g_i^{\ k} F_{mn} F^{mn}\rt] ,
\ee
\be
\frac{1}{\sqrt{-g}} \lt(\sqrt{-g} F^{ik}\rt)_{,k} = 4 \pi J^i ,
\ee
\be
F_{ij,k}+F_{jk,i}+F_{ki,j} = 0 \ .
\ee
$R_i^{\ k}$ is the Ricci tensor.
$T_i^{\ k}$ and $E_i^{\ k}$ are the energy-momentum tensors due
to the matter and the electromagnetic field, respectively. $J^i$ stands
for the electric current density vector.

\begin{flushleft}
{\large {\bf 3. Solution for charged tachyon}}\\
\end{flushleft}

By transforming the Reissner-Nordstr\"{o}m solution, we
have obtained  a static and axially symmetric exact solution of the
Einstein-Maxwell equations which is given by the line element,
\be
d\tau^2\  =\  B dt^2 - B^{-1} d\rho^2 - \frac{\rho^2}{(1-v\ cos\th)^2}\
                         \lt(d\th^2 + sin^2\th\ {d\phi}^2\rt)\ ,
\ee
where
\be
B\ =\  1 - v^2 + \frac{m}{\rho} +\frac{Q^2}{\rho^2} {\lt(1-v^2\rt)}^2
\ee
and the only component of the electromagnetic field tensor
\be
F_{\rho t}\ =\ -\ \frac{Q\ (1-v^2)}{\rho^2} \ .
\ee
The non-vanishing components of the energy-momentum tensor of
the electromagnetic field are
\be
E_t^{\  t}\ = \ E_{\rho}^{\  \rho}\  =\ - E_{\th}^{\  \th}\ =\ -E_{\phi}^{\
\phi}\
=\  \frac{Q^2}{8\pi \ {\rho}^4}\ {\lt(1-v^2\rt)}^2 \ .
\ee
The current density  vector and the energy-momentum tensor of
matter are given by
\be
J^i\ =\  0, \nn\\
\ T_i^{\  k}\ =\ 0 \ .
\ee
We now proceed to show that the above solution gives the
gravitational and electromagnetic fields of a charged tachyon.

Using the retarded  time coordinate $u$ given by
\be
u\  =\ t\ -\ \int \lt[B(\rho)\rt]^{-1}  \  d\rho
\ee
 in place of t, the above line element  can
be written as
\be
d\tau^2\ =\ B du^2 + 2 du d\rho - \frac{{\rho}^2}{(1-v\ cos\th)^2}
\lt(d\th^2 + sin^2\th \ d\phi^2\rt)
\ee
and the non-vanishing component of the electromagnetic field tensor is
given by
\be
F_{\rho u}\ =\ -\ \frac{Q\ (1-v^2)}{\rho^2} \ .
\ee
Now by a coordinate transformation $\rho$ going to $r$, where
$\rho = r (1- v \ cos\th)$, the line element is
\be
d\tau^2\ =\ D du^2 + 2 du\  d\lt[r(1-v\ cos\th)\rt]
         - r^2 (d{\th}^2 + sin^2 \th \ d\ph^2) \  ,
\ee
where
\be
D\ =\ 1 - v^2 + \frac{m}{r(1-v\ cos\th)}
              + \frac{Q^2(1-v^2)^2}{r^2(1-v\ cos\th)^2}
\ee
and
the surviving components of the electromagnetic field tensor are
\bea
F_{ru}&=&-\ \frac{Q (1-v^2)}{r^2 \ (1-v\ cos\th)}\nn\\  \ ,
F_{{\th}u}&=&-\ \frac{Q (1-v^2)\ v\  sin\th}{r\ {(1-v\ cos\th)}^2}
\eea
For $|v|<1$,  the  solution can be transformed to the
Reissner-Nordstr\"{o}m solution given by the line element,
\be
d\tau^2\ =\ B'\ d{u'}^2 +\ 2 du'\ dr'
- {r'}^2 \lt(d{\th'}^2 + sin^2{\th'}\ d\phi^2\rt)\ ,
\ee
where
\be
B'\ =\ 1 - \frac{2M}{r'} + \frac{Q^2}{{r'}^2}
\ee
and
\be
M\ =\ -\  \frac{m\ (1-v^2)^{-3/2}}{2} \ ,
\ee
and the non-vanishing component of the electromagnetic field tensor
\be
F_{r'u'}\  =\ - \frac{Q}{{r'}^2} \ ,
\ee
by the  coordinate transformations
\bea
u'&=&u \sqrt{1-v^2}\ ,\nn\\
r'&=&\frac{{r} (1 - v\ cos\th)}{\sqrt{1 - v^2}}\ ,\nn\\
cos\th'&=&\frac{cos\th - v}{1 - v\ cos\th} \ .
\eea
We obtained the axially symmetric solution by following the reverse process
given here, i.e. we began with the Reissner-Nordstr\"{o}m solution and
 obtained
the axially symmetric solution by a complex transformation ($|v|>1$). However,
for convenience in presentation we have first written the axially symmetric
solution ( the line element  in diagonal form) and then have shown that
it transforms to the  Reissner-Nordstr\"{o}m solution for $|v|<1$.
The coordinate transformations expressed  by
($20$) show that the origin of the frame $S'$  moves with respect to the
frame $S$ with a uniform velocity $v$ along the common Z-axis in the flat
 spacetime background $[12]$. The solution of the Einstein-Maxwell
equations obtained by us  has the following characteristics:
(a) the gravitational field given by the line element ($5$) has
all the geometrical characteristics of the Reissner-Nordstr\"{o}m field,
(b) the solution is axially symmetric which is required as
there cannot be a tardyonic observer which can find a tachyon at
rest, and  (c) it contains a velocity parameter $v$,  which is not
possible to be transformed away for $|v|>1$.  However, for $|v|<1$,
the solution transforms to the Reisnner-Nordstr\"{o}m solution.
Therefore, we interprete the solution to give the gravitational
and the electromagnetic fields of a non-rotating charged tachyon.
 Switching off the charge parameter Q,  one gets the solution for the
uncharged (non-rotating) tachyon which was obtained by Vaidya $[12]$

\begin{flushleft}
{\large {\bf 4. Discussion}}\\
\end{flushleft}

The solution given by us is singular at $r=0$ and
$sec\th=v$. The latter gives a right circular cone of semi-vertical
angle $arcsec(v)$ as the singularity surface. Again the null surfaces are
given by the equation
\be
(v^2-1)\ {\rho}^2\ -\ m \rho\ -\ Q^2\ (v^2-1)^2\ =\ 0 \ .
\ee
This equation has two roots given by
\be
\rho\ =\ {\rho}_{\pm} \ ,
\ee
where
\bea
\rho_{+}&=&\frac{m+\sqrt{m^2 + 4 (v^2-1)^3 Q^2}}
                  {2 (v^2-1)},\nn\\
\rho_{-}&=&\frac{m-\sqrt{m^2 + 4 (v^2-1)^3 Q^2}}
                  {2 (v^2-1)} \ .
\eea
Using $\rho = r (1- v cos\th)$, $ r = (x^2 + y^2 + z^2)^{1/2} $,
and $r\ cos\th = z$, the equations of null surfaces given by (22) can be
written as
\be
\lt(z \sqrt{v^2-1} + \frac {v\ {\rho}_{\pm} }{\sqrt{v^2-1}}\rt)^2\
-\ (x^2 + y^2)\ =\ \frac{ {{\rho}_{\pm}}^2 } {v^2-1}
\ee
which give two hyperboloids of revolution of two sheets in the
three (spatial) space corresponding to $\rho$ equals to
$\rho_{+}$ and $\rho$ equals to $\rho_{-}$. It is clear from $(22)-(23)$
that the null surfaces for the charged tachyon are different from
those of the uncharged tachyon. It is of  interest to study the
trajectories of the charged  tardyons in the field of a charged tachyon.

After  the completion of this work, a  paper $[15]$  came to
our notice wherein the authors, using the complex tetrads,  obtained
a general Kerr-Schild class of solutions. They  mentioned
that a particular solution   gives the charged rotating
tachyon. The solution given by us  has a clear physical interpretation.
 Dadhich brought to our notice his paper
on charged tachyon [16] .
Our method of obtaining the gravitational and electromagnetic fields
of a charged tachyon is different from his.

\begin{flushleft}
{\bf Acknowledgements}\\
\end{flushleft}
This work  is  supported in part by the Basque government.
Thanks are due to  P. C. Vaidya for his guidance throughout the
preparation of this work, and to  J. C. Parikh and
Alberto Chamorro  for discussions.

\newpage
\parindent 0pt
\begin{flushleft}
{\Large{ References }}\\
\end{flushleft}
\hsize 8in

$[1]$\  G  Fienberg, {\it Phys. Rev.}  \ {\bf 159} \ 1089\ (1967).\\
$[2]$\ O M P Bilaniuk and E. C. G. Sudarshan, {\it Phys. Today}  \
{\bf 5} \ 43\ (1969).\\
$[3]$\  E. Recami, {\it Riv. del Nuovo Cim.} \ {\bf 9} \ 1\ (1986).\\
$[4]$\ O M P Bilaniuk  V. K. Deshpande  and E. C. G. Sudarshan,
{\it Am. J. Phys.} \ {\bf 30} \\
{}~~718\ (1962).\\
$[5]$\ T Alv\"{a}ger and M. N. Kreisler, {\it Phys. Rev.} \ {\bf 171} \ 1357\
(1968).\\
$[6]$\ M V Davis  T. Alv\"{a}ger and M. N. Kreisler, {\it Phys. Rev.} \ {\bf
183} \ 1132\ (1969).\\
$[7]$\ D F Bartlett and M D Lahana, {\it Phys. Rev. D}  \ {\bf 6} \
1817\ (1972).\\
$[8]$\ P V Ramana Murthy, {\it Lett. Nuov. Cim.}  {\bf 1} \ 908\ (1971).\\
$[9]$\ J V Narlikar and E C G Sudarshan, {\it Mon. Not. R. astr.
Soc.}  {\bf 175} \ 105\ (1976).\\
$[10]$\ J V Narlikar and S V Dhurandhar, {\it Pramana-J. Phys.} \ {\bf
6} \  388\ (1976).\\
$[11]$\ S V Dhurandhar, {\it Phys. Rev.}  {\bf 19} \ 2310\ (1979).\\
$[12]$\ P C Vaidya, {\it Curr. Sci. (India)}  \ {\bf 40 } \  651\ (1971).\\
$[13]$\ A K Raychaudhuri, {\it J. Math. Phys.}  {\bf 15} \ 856\ (1974).\\
$[14]$\ K S Virbhadra, {\it Pramana-J. Phys.} {\bf 40} \ 273 \ (1993)\\
$[15]$\ P V Bhatt and S K Vaidya, {\it Class. Quant. Gravit.} \ {\bf
8} \ 1717\ (1991).\\
$[16]$\ N Dadhich, {\it Indian J. Pure} \& {\it Appl. Math.}  {\bf 7}  2 \ 151\
 (1976).
\end{document}